\documentclass[final]{WileyASNA-v1}

\articletype{Article Type}%

\received{...}
\revised{...}
\accepted{...}

\begin{document}

\title{1321+045: a Compact Steep Spectrum radio source in a cool-core galaxy cluster}

\author[1]{Ewan O'Sullivan*}

\author[2]{Magdalena Kunert-Bajraszewska}

\author[1]{Aneta Siemiginowska}

\author[1]{D.J. Burke}

\author[3,4]{Fran\c{c}oise Combes}

\author[3]{Philippe Salom\'{e}}

\author[5]{Simona Giacintucci}

\authormark{Ewan O'Sullivan \textsc{et al}}

\address[1]{\orgname{Center for Astrophysics $|$ Harvard \& Smithsonian}, \orgaddress{60 Garden Street, Cambridge \state{MA} 02138, \country{USA}}}

\address[2]{\orgdiv{Institute of Astronomy}, \orgname{Faculty of Physics, Astronomy and Informatics, NCU}, \orgaddress{Grudzi\k{a}dzka 5, 87-100 Toru\'{n}, \country{Poland}}}

\address[3]{\orgdiv{LERMA, Observatoire de Paris}, \orgname{CNRS, PSL Univ., Sorbonne Univ.}, \orgaddress{61 Avenue de l'Observatoire, 75014 Paris, \country{France}}}

\address[4]{\orgname{Coll\`{e}ge de France}, \orgaddress{11 Place Mercelin Berthelot, 75005 Paris, \country{France}}}

\address[5]{\orgname{Naval Research Laboratory}, \orgaddress{4555 Overlook Avenue Sw, Code 7213, Washington, \state{DC} 20375, \country{USA}}}

\corres{*Ewan O'Sullivan, 60 Garden Street, Cambridge, MA 02138, USA. \email{eosullivan@cfa.harvard.edu}}

\abstract{Cluster-central gigahertz peak and compact steep spectrum (CSS) sources offer an opportunity to study the earliest phases of AGN feedback, but few have yet been examined in detail. We present results from radio and X-ray observations of 1321+045, a CSS source in a 4.4~keV cluster at $z$=0.263. The cluster has a strongly cooling core, and disturbances from a minor cluster merger may have triggered a period of jet activity which formed the 16~kpc radio lobes 2.0$^{+0.3}_{-0.2}$~Myr ago. However, new VLBA imaging shows a $\sim$20~pc jet on a different projected axis, which is probably only a few hundred years old. We consider possible histories for the system, with either one or two periods of jet activity. While this single system is informative, a broader study of the youngest cluster-central radio sources is desirable.}


\jnlcitation{\cname{%
\author{O'Sullivan E.}, 
\author{M. Kunert-Bajraszewska}, 
\author{A. Siemiginowska}, 
\author{D.J. Burke}, 
\author{F. Combes}, 
\author{P. Salom\'{e}}, and
\author{S. Giacintucci}} (\cyear{2021}), 
\ctitle{1321+045: a Compact Steep Spectrum source in a cool-core galaxy cluster}, \cjournal{A.N.}, \cvol{2021;00:1--4}.}

\fundingInfo{National Aeronautics and Space Administration (NASA) \textit{Chandra} Award Number GO0-21112X; National Science Centre, Poland, grant no. 2017/26/E/ST9/00216; Naval Research Laboratory 6.1 Base funding}

\maketitle

\section{Introduction}\label{sec:intro}

It is now well established that the dominant galaxies of galaxy clusters are around an order of magnitude more likely to host radio AGN than non-central galaxies of equivalent mass \citep{Bestetal07}. Their position at the centres of the most massive dark matter haloes in the Universe means that they host the most massive black holes \citep{Bogdanetal18,Gasparietal19}, while cooling from a relaxed intra-cluster medium (ICM) can supply the gas needed to fuel repeated periods of jet activity \citep[e.g.,][]{Babyketal19}. Observations of the ICM provide an  important window on these central radio sources, tracing not only the cooling processes by which they are fuelled, but, via detection of shocks and cavities associated with expanding radio lobes, providing accurate measurements of the mechanical power of the radio jets \citep[e.g.,][]{OSullivanetal11b}, and information on their particle content. Conversely, the properties of the central radio source show where any given cluster may be in its feedback cycle, from compact young sources just beginning their outburst, through full-size FR~I \citep[or more rarely FR~II,][]{FanaroffRiley74} galaxies with jets actively heating the ICM, through to steep-spectrum sources passively aging after their jets have shut down.

Clusters hosting the youngest radio sources may be of particular interest if we wish to examine the ICM conditions which trigger AGN feedback. Gigahertz peak spectrum (GPS) and compact steep spectrum (CSS) sources are believed to be among the youngest radio galaxies, typically $<$10$^5$~yr old \citep{Fantietal95,Readheadetal96b} and, at least in some cases, at the start of development into much larger plumed or lobed FR~Is or IIs \citep{Readheadetal96b}. 

\begin{figure*}[t]
\centerline{
\includegraphics[height=6cm,viewport=10 37 310 335,clip=]{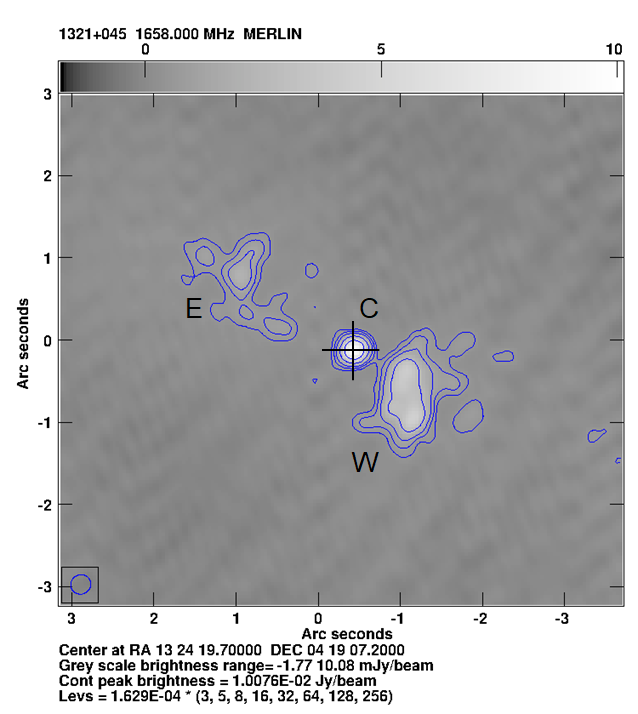}
\includegraphics[height=6cm,viewport=10 37 305 340,clip=]{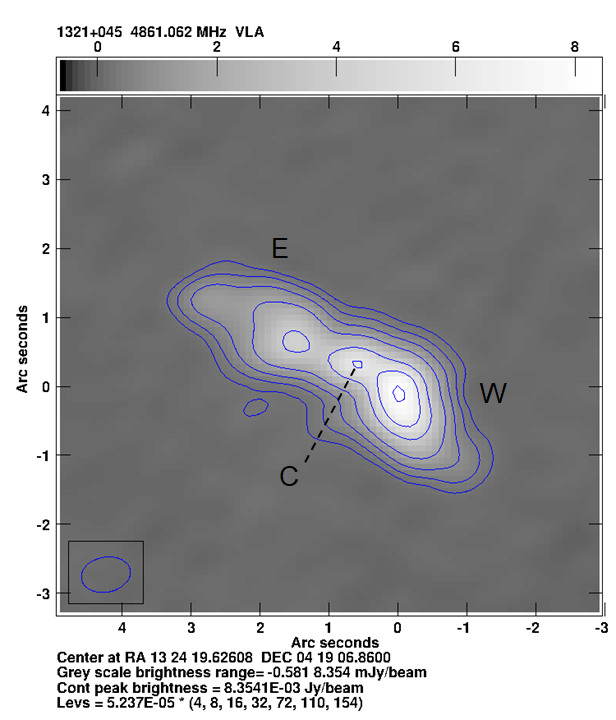}
\includegraphics[height=6cm,viewport=6 32 260 343,clip=]{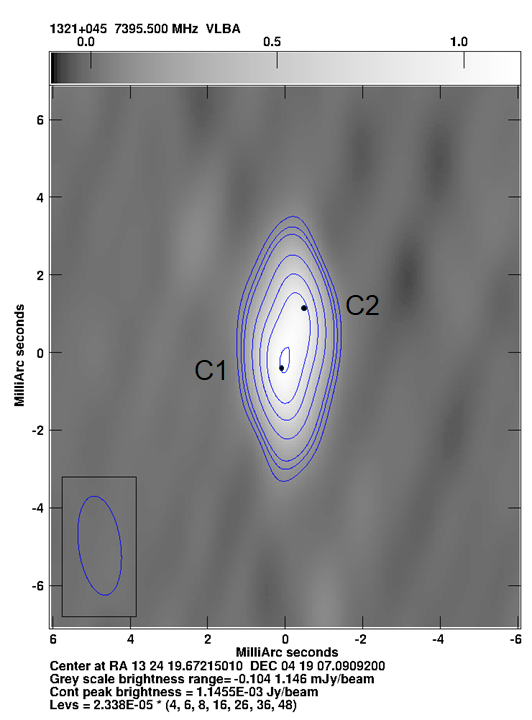}
}
\caption{\label{fig:radio}From left to right, MERLIN 1.6~GHz, VLA 4.9~GHz and VLBA~7.5~GHz images of 1321+045. The colour bars indicate the flux density in mJy~beam$^{-1}$. The beam size and r.m.s. noise levels are 0.25$^{\prime\prime}$$\times$0.24$^{\prime\prime}$ and 163~$\mu$Jy, 0.72$^{\prime\prime}$$\times$0.51$^{\prime\prime}$ and 52~$\mu$Jy, and 2.58mas$\times$1.03mas and 23$\mu$Jy, respectively.}
\end{figure*}

Unfortunately, relatively few cluster central GPS or CSS sources have been studied in detail. Spectral studies of cluster-central sources have shown that $\sim$8.5\% contain GPS-like features \citep{Hoganetal15b}, e.g., spectra which follow a power law at low frequency but which invert and show a peak at high frequencies. Some of these may be restarted sources, where the low-frequency component comes from old lobes, but others may be examples of jets which are continuous but whose power is variable. Confirmation of a new phase of jet activity can be provided by imaging studies, as in the nearby group-dominant galaxy NGC~5044, where X-ray and radio observations have shown three epochs of old jet activity, but recent VLBA data show that a new jet has been launched on a different axis \citep{Schellenbergeretal21}. Examples of cluster-central CSS sources include the quasars 3C~186 \citep{Siemiginowskaetal05,Siemiginowskaetal10,Migliorietal12} and IRAS~F15307+3252 \citep{HlavacekLarrondoetal17}. The former is located in an 8~keV cluster at $z$=1.06, but the AGN is so bright in X-rays that detailed analysis of the cooling region is impossible, while the latter occupies a 2~keV group at $z$=0.93 and is faint enough that only global properties can be determined using current X-ray observatories.


A third cluster-central CSS radio galaxy is 1321+045, in the $z$=0.263 cluster MaxBCG~J201.08197+04.31863. The radio source has a projected size $\sim$4$^{\prime\prime}$ ($\sim$16~kpc) with an FR~I morphology and a clearly separated core \citep{Kunert-Bajraszewskaetal10}. A 9~ks \textit{Chandra} snapshot showed the surrounding cluster to have a temperature $\sim$4.4~keV and a cool core, and suggested the radio lobes were over-pressured by a factor $\sim$2 \citep{Kunert-Bajraszewskaetal13}. The cluster galaxy population appears relatively relaxed \citep{WenHan13} and the brightest cluster galaxy (BCG)  has an H$\alpha$ luminosity comparable to those of the cooling nebulae seen around low-$z$ BCGs \citep[L$_{\rm H\alpha}$=4.5$\times$10$^{41}$~erg~s$^{-1}$][]{Liuetal12}. As a reasonably well-resolved source in a luminous cluster, with no bright X-ray core to contaminate observations, 1321+045 provides an excellent opportunity to study the conditions that may trigger a cluster-central AGN.

Throughout the paper we adopt a flat cosmology with H$_0$=70~km~s$^{-1}$~Mpc$^{-1}$, $\Omega_\Lambda$=0.7 and $\Omega_{\rm M}$=0.3. For the BCG redshift of $z$=0.263, this gives an angular scale of 1$^{\prime\prime}$=4.058~kpc.

\section{X-ray and Radio observations of 1321+045}
 We acquired new observations of 1321+045, including a $\sim$80~ks \textit{Chandra} exposure, a 2.5~hr VLBA C-band integration, and 12~hr IRAM~30m CO(1-0) and CO(3-2) observations. The results are reported in detail in \citet{OSullivanetal21a} and summarized here. The CO observations failed to detect the BCG, providing only upper limits of M$_{\rm mol}\leq$7.7$\times$10$^9$~M$_\odot$ and M$_{\rm mol}\leq$5.6$\times$10$^9$~M$_\odot$ from the CO(1-0) and CO(3-2) transitions respectively. However, the radio and X-ray observations reveal important new information about the system.

Figure~\ref{fig:radio} shows radio continuum images of the source from the Multi-Element Radio Linked Interferometer Network (MERLIN) at 1.6~GHz, Very Large Array (VLA) at 4.9~GHz, and Very Long Baseline Array (VLBA) at 7.5~GHz. The VLA image confirms the same structure as found by \citet{Kunert-Bajraszewskaetal10} from the MERLIN data, with east and west lobes (E and W) and a core (C) coincident with the BCG optical centroid. Modelling of the 74~MHz - 4.9~GHz spectrum of the source (including lobes and core) shows a break at 147$^{+39}_{-36}$~MHz, which for the measured equipartition magnetic field strength of the source, $\sim$150~$\mu$G, implies a lobe age of 2.0$^{+0.3}_{-0.2}$~Myr. The spectrum shows no sign of the high-frequency break expected if the jets no longer power the lobes, implying that if they have ceased to do so, it is only in the last $<$0.11~Myr. The VLBA image probes much smaller angular scales and reveals a $\sim$5~mas ($\sim$20~pc) extension of the core component. We interpret this as a jet, with the difference between its projected axis and that of the radio lobes suggesting that it may be newly launched. If so, it is probably only a few hundred years old, based on typical expansion times of compact radio sources. 

\begin{figure}[t]
\includegraphics[width=0.99\columnwidth,bb=0 0 660 660]{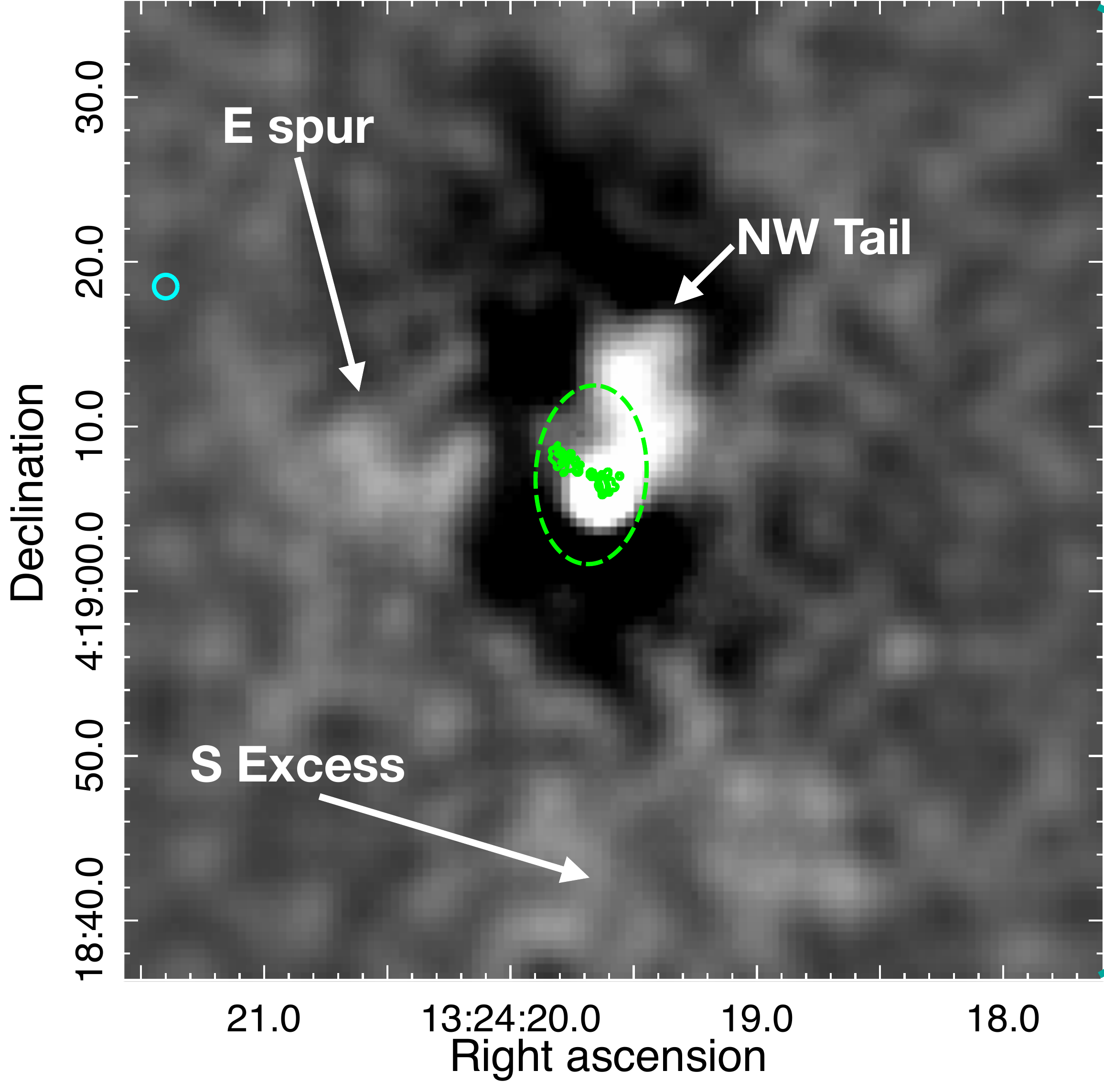}
\caption{\label{fig:resid}\textit{Chandra} 0.5-7~keV surface brightness residual map, after subtraction of a model of the ICM. MERLIN 1.6~GHz contours are shown in green, the dashed ellipse marks the approximate D$_{25}$ contour of the BCG, and the cyan circle the position of the second-ranked galaxy. Disturbed ICM structures are labelled.}
\end{figure}

Analysis of the \textit{Chandra} data shows that the cluster has likely been disturbed by a recent minor merger. Subtraction of a surface brightness model describing the large-scale ICM reveals a small tail of gas extending $\sim$10$^{\prime\prime}$ ($\sim$40~kpc) to the northwest from the BCG, and larger-scale features to the south and east of the cool core (see Figure~\ref{fig:resid}). Maps of ICM temperature and entropy show that the southern feature is a tail of cool, low-entropy gas extending at least 45$^{\prime\prime}$ (180~kpc), while the eastern spur also consists of cool gas and extends toward the second brightest galaxy in the cluster. Such structures are often formed by the infall of a low-mass cluster or galaxy group. As the smaller system falls in, ram-pressure forces strip its hot gas halo away from the galaxies, leaving behind a tail of cooler material. The apparent link with the second-ranked galaxy, SDSS J132421.40+041918.5, may indicate that this was the BCG of the infalling system.

Radial profiles of ICM properties show that the cluster is comparable to strongly cooling systems in the local Universe, but has particularly low central entropy and cooling time (8.6$^{+2.2}_{-1.4}$~keV~cm$^2$ and 390$^{+170}_{-150}$~Myr within 8~kpc). The ratios of cooling time to free-fall time (t$_{\rm cool}$/t$_{\rm ff}$) and eddy turnover time (t$_{\rm cool}$/t$_{\rm eddy}$) suggest that the ICM becomes thermally unstable within a radius of $\sim$45~kpc; within this region, ICM gas is likely to cool and condense to form dense molecular clouds which can flow into the BCG to fuel the AGN. The overall X-ray luminosity of the cooling region (within which t$_{\rm cool}<$7.7~Gyr) is $\sim$3.1$\times$10$^{44}$~erg~s$^{-1}$, similar to the estimated jet power, $\sim$1.4$\times$10$^{44}$~erg~s$^{-1}$, suggesting the AGN is capable of maintaining thermal balance if its jet activity continues.

\section{History of jet activity}
While the strong cool core of the cluster provides the conditions necessary to fuel the AGN, the presence of a small-scale tail of gas associated with the BCG suggests that the cluster merger has disturbed the ICM on scales all the way down to the central few kiloparsec. This disturbance may have triggered jet launching. However, the mismatch between the projected axes of the radio lobes and VLBA jet makes the recent history of the AGN unclear.

One possibility is that we are observing two eras of jet activity. The first, perhaps triggered by the minor merger, occurred $\sim$2~Myr ago and produced the radio lobes. Then, within the last 0.11~Myr, the radio jets shut down and the AGN reoriented, with a second pair of jets being launched on a new axis a few hundred years ago. The difficulty with this scenario comes from the very short timescale in which reorientation must occur. Such a reorientation could be caused by a merger with another supermassive black hole (SMBH), or torques applied by a massive thin accretion disk during a period of high accretion rates, but we see no evidence that the BCG has recently merged with another galaxy, and the AGN shows no indication of the broad ionisation lines, high optical and X-ray luminosity associated with rapid accretion. Accretion at the low rates typical for radiatively inefficient, jet-launching AGN would take far too long to alter the spin axis of the SMBH. 

An alternative is that the jets are aligned close to the line of sight. A small difference in \textit{true} jet axis might then produce the observed large difference in \textit{projected} axis. Precession of the jet could have been caused by relativistic frame dragging of a misaligned accretion disk \citep[e.g.,][]{McKinneyetal13,Liskaetal18} even if accretion rates are low. The difference in brightness of the two lobes, and the fact that the VLBA data show only a one-sided jet, could be consistent with an axis near the line of sight, with Doppler boosting brightening the approaching jet and dimming its receding counterpart to the point where it cannot be detected in our observation. However, the angle to the line of sight would need to be very small, and even taking into account a degree of jet bending, we would expect the accretion disk to be face-on to us, with little obscuration. Thus the lack of a clear AGN signature in optical or X-ray bands is again problematic.

\section{Conclusions}\label{sec:Conc}
Cluster-central CSS and GPS radio sources offer an opportunity to study the conditions which trigger AGN feedback and the impact on the surrounding ICM during its early phases, as well as providing additional information on the properties of the radio sources themselves. Since the BCGs of cool-core galaxy clusters are known to undergo repeated cycles of jet activity fuelled by cooling from the ICM, we should expect their AGN to go through GPS and CSS states in each cycle. However, since these phases are short-lived, and the presence of old, steep-spectrum radio structures from previous outbursts may complicate identification, it is perhaps unsurprising that to date only a handful of such systems have been studied.

Combining X-ray and radio observations of the $z$=0.263 cluster-central CSS 1321+045, we have shown that its cluster has properties comparable to the most rapidly cooling strong cool-core clusters in the local Universe, and that disturbances probably associated with a recent minor cluster merger may have triggered the fuelling and outburst of its AGN. However, the history of the AGN jets is still uncertain, and further high spatial-resolution radio observations will be necessary before we can ascertain whether the difference in projected axis between the VLBA jet and the outer lobes is the product of multiple periods of jet activity, or precession of jets close to the line of sight.

While this study of a single system has been highly informative, a sample of young cluster-central sources is obviously required if we are to gain a clear picture of the early stages of AGN feedback. To date, CSS and GPS sources have typically been identified from radio surveys \citep[e.g.,][]{Kunert-Bajraszewskaetal10,Wolowskaetal21}, with environment investigated only for subsets \citep[e.g.,][]{Siemiginowskaetal16}. Given the significant fraction ($\sim$8.5\%) of BCGs with GPS-like spectra, perhaps the time has now come to for a targeted study of young radio sources in cluster-dominant galaxies.

\bibliography{../paper}

\section*{Author Biography}

\begin{biography}{}{\textbf{Ewan O'Sullivan} is an Astrophysicist working in the \textit{Chandra} X-ray Center at the Center for Astrophysics in Cambridge, MA, USA. His research interests include the X-ray properties of galaxy groups, clusters, and individual ellipticals, AGN feedback in those systems, and the formation and evolution of galaxy groups.}
\end{biography}

\end{document}